\documentclass[12pt,a4paper]{article}
\usepackage{authblk}
\usepackage{a4wide}	
\usepackage{amssymb,mathtools,amsmath,amsthm,bm}
\usepackage{graphicx}
\usepackage{hyperref}
\usepackage[utf8]{inputenc}
\usepackage{babel}
\usepackage[dvipsnames]{xcolor}
\usepackage{caption}
\usepackage{subcaption}

\def\p{{\partial}}
\def\Re{\mathop{\text{Re}}}

\usepackage{xcolor}

\newmuskip\pFqmuskip

\newcommand*\pFq[6][8]{%
  \begingroup 
  \pFqmuskip=#1mu\relax
  \mathchardef\normalcomma=\mathcode`,
  \mathcode`\,=\string"8000
  \begingroup\lccode`\~=`\,
  \lowercase{\endgroup\let~}\pFqcomma
  {}_{#2}F_{#3}{\left(\genfrac..{0pt}{}{#4}{#5};#6\right)}%
  \endgroup
}
\newcommand{\pFqcomma}{{\normalcomma}\mskip\pFqmuskip}%

\title{Universality of stochastic Laplacian growth}
\author[1]{Oleg Alekseev}
\affil[1]{Chebyshev Laboratory, Department of Mathematics and Computer Science, Saint-Petersburg State University, 14th Line, 29b, 199178, Saint-Petersburg, Russia}
\date{}
\begin{document}
\maketitle
\begin{abstract}
We consider a stochastic Laplacian growth problem in the framework of normal random matrices. In the large $N$ limit the support of eigenvalues of random matrices is a planar domain with a sharp boundary which evolves under a change in the size of matrices. This evolution can be interpreted as a stochastic growth process. We show that the most probable growth scenario is similar to deterministic Laplacian growth. The other scenarios involve Laplacian growth in the presence of fluctuations. We use the random matrix approach to determine a probability distribution function of fluctuations. The partition function of fluctuations is shown to be universal. It does not depend on the shape of the initial domain and depends on the strength of fluctuations and the geometry of the problem only.
\end{abstract}


\section{Introduction}


Laplacian growth (LG) is a fundamental problem of pattern formation in nonequilibrium physics~\cite{Cross93, Gollub99}. A distinctive feature of the problem is the instability of the interface dynamics. A relevant example is a Hele-Shaw flow when the inviscous fluid (water) is injected into the viscous fluid (oil) in a narrow gap between two parallel plates, known as a Hele-Shaw cell~\cite{BensimonRMP, SaffmanTaylor}. When the effect of surface tension is negligibly small, the problem possesses a reach integrable structure unusual for most nonlinear physical phenomena~\cite{Bensimon84}. It is fully integrable, has infinitely many conservation laws~\cite{Richardson}, and is equivalent to the dispersionless 2D Toda hierarchy~\cite{WiegmannPRL}. The integrable structure of LG is identical to the one observed in the models of normal random matrices~\cite{BOOKtau}.

Another relevant example of LG phenomena is the semiclassical behavior of electronic droplets in the quantum Hall regime. The droplet of electrons can be considered as a support of eigenvalues of normal random matrices in the semiclassical limit~\cite{Chau92}. As one changes the degeneracy of the level, the number of electrons in the droplet increases. Remarkably, the growth probability of the droplet is determined by the harmonic measure of its boundary~\cite{ABWZ02}. This growth law coincides with the definition of diffusion-limited aggregation (DLA)~\cite{WittenSanderPRL}, which is a discrete universal stochastic fractal growth of the Laplacian kind.

The interpretation of normal random matrices in terms of an aggregation process seems to be a productive approach in several applications. One of them is stochastic Laplacian growth, which is developed as a bridge between the deterministic LG and DLA~\cite{AM16}. In this paper, we show that stochastic LG can be naturally embedded in the framework of normal random matrices. As a result, we obtain a probability distribution of random fluctuations at the boundary of the growing droplet. Remarkably, the partition function of fluctuations is shown to be universal. It only depends on the geometry of the problem where the interface dynamics is considered (e.g., LG on the plane, in an infinite rectangular channel, or a wedge) in a simple manner.

This paper is organized as follows: in the next section, we review the ensemble of normal random matrices to be considered and give their physical interpretations. In Section~\ref{s:growth} we discuss applications of random matrices to the growth processes and the discrete stochastic Laplacian growth model~\cite{AM16}. We study the semiclassical limit of the growth process of the support of eigenvalues and show that the most probable (classical) dynamics of the boundary prompts Darcy's law. It is worth mentioning Section~\ref{s:universality} where we show that the partition function of the fluctuations is universal. The Appendices contain technical details of some proofs and calculations.


\section{Partition function of normal random matrices}\label{s:model}


We start this section with a brief review of the ensemble of normal random matrices. The matrix $\mathsf M$ is called normal if it commutes with its Hermitian conjugation $[\mathsf M,\mathsf M^\dagger] = 0$. This means that both matrices, $\mathsf M$ and $\mathsf M^\dagger$, can be diagonalized simultaneously. Namely, let $\mathsf M = \mathsf U \mathsf Z \mathsf U^\dagger$, where $\mathsf U$ is a unitary matrix and $\mathsf Z= diag(z_1,\dotsc,z_N)$ is the diagonal matrix of $N$ complex eigenvalues of $\mathsf M$. In terms of eigenvalues the partition function can be written as the following $N$-fold integral over the complex plane~\cite{Chau92}:
\begin{equation}\label{tau-def}
	\tau_N = \frac{1}{N!}\int_{\mathbb C}\cdots \int_{\mathbb C}|\Delta_N(z)|^2\prod_{j=1}^N e^{\frac{1}{\hbar}W(z_j,\bar z_j)}d^2z_j.
\end{equation}
Here $\Delta_N(z) = \prod^N_{m<n} (z_m-z_n)$ is the Vandermonde determinant, $d^2z: = dx dy$ is the area element, $\hbar$ is a parameter of the model, and $W(z,\bar z)$ is the so-called external potential. Below, we consider the following family of potentials:
\begin{equation}\label{W-def}
	W(z,\bar z) = -U(z,\bar z) + \sum_{k\geq1}(t_kz^k + \bar t_k \bar z^k),
\end{equation}
where the parameters $\{t_k,\bar t_k\}$ are the coupling constants of the matrix model, and $U(z,\bar z)$ is a non-harmonic part of the potential. We assume that $U(z,\bar z)$ is a real-analytic and real-valued function in $\mathbb C\setminus\{0\}$. It is convenient to introduce the function 
\begin{equation}\label{sigma-def}
	\sigma(z,\bar z) = \p\bar\p U(z,\bar z)>0,
\end{equation}
where $\p=\p/\p z,\ \bar\p = \p/\p\bar z$. We refer to $\sigma$ as the background charge density in the complex plane, and $U(z,\bar z)$ is the electrostatic potential created by these charges.

The integral~\eqref{tau-def} has a clear physical interpretation. Namely, it represents the partition function of the 2D statistical ensemble of $N$ Coulomb charges in the external potential $W$~\cite{DysonJMP1}. Let us examine the so-called \textit{semiclassical} limit of the matrix model, which is a large $N$ limit where $\hbar $ approaches zero while keeping $N\hbar$ fixed. In this scenario, the eigenvalues of matrices from the ensemble densely occupy a connected domain $D$ in the complex plane with a sharp edge. This domain is referred to as the support of eigenvalues or \textit{droplet}. The eigenvalues are distributed with density~\eqref{sigma-def} inside the droplet. The shape of the droplet is determined by solving the inverse potential problem. In the case of potential~\eqref{W-def} the domain $D$ is determined by the relations:
\begin{equation}\label{tk-def}
	t_0=\frac{1}{\pi}\int_D \sigma(z,\bar z)d^2z,\quad t_k = -\frac{1}{\pi k}\int_{D^c}\sigma(z,\bar z)z^{-k}d^2z,\quad k\geq1,
\end{equation}
where $D^c = \mathbb C\setminus D$ is a complement to the support of eigenvalues. These equations relate the shape of $D$ with the parameters of the external potential $W$. Note, that $\pi t_0 = N\hbar$ corresponds to the area of the droplet, and $-k t_k$ are equal to the harmonic moments of $D^c$. The integral representation for $t_0$ implies that $\sigma$ is integrable everywhere in $D$. If $\sigma$ is singular at some point, one has to introduce a cutoff~\cite{Z13}. One can prove that $t_0$ together with the parameters $\{t_k,\bar t_k\}$ form a set of local coordinates in the space of domains~\cite{NZ15}. Hence, one can identify functional on the space of planar domains $D$ with a function of infinitely many independent variables $t_0$, $\{t_k, \bar t_k\}$.

The semiclassical limit, $N\hbar=const$, implies that the conventional $1/N$-expansion of the matrix model~\eqref{tau-def} is equivalent to the $\hbar$-expansion,
\begin{equation}\label{tau-asymptotic}
	\tau_N = (c_p\hbar)^{N/2}\exp\left(\frac{F_0}{\hbar^2}+\frac{F_{1/2}}{\hbar}+F_1+O(\hbar)
    )\right),
\end{equation}
where $c_p$ is a numerical constant, which depends on the non-harmonic part of the potential $U(z,\bar z)$. For quasi-harmonic potentials with $U(z,\bar z)=|z|^2$, the constant can be found explicitly $c_{p} = 2\pi^3$. The leading contribution to the free energy, $F\equiv F_0 = \lim_{\hbar\to0} \hbar^2 \log \tau$, is a functional on the domain $D$:
\begin{equation}\label{F0}
	F(t_0, \{t_j,\bar t_j\})= -\frac{1}{\pi^2}\iint_{D}\sigma(z,\bar z)\log\left|z^{-1}-\zeta^{-1}\right| \sigma(\zeta,\bar \zeta)d^2z d^2\zeta.
\end{equation}
Remarkably, $F$ is the dispersionless tau-function of the 2D Toda hierarchy~\cite{BOOKtau, Z01}. It obeys the dispersionless Hirota relations and possesses remarkable properties briefly described below.

Below, we need a univalent conformal map from the exterior of the unit disk $\mathbb D^c$ onto the domain $D^c$. Let us introduce  the conformal map $z(w)$ from $\mathbb D^c$ onto $D^c$ normalized so that $z(\infty) = \infty$ and the conformal radius $r = \lim_{w\to\infty} z(w)/w$ is real. In general, the Laurent expansion of the function $z(w)$ around infinity is
\begin{equation}\label{z-def}
	z(w) = rw + \sum_{k\geq0}u_k w^{-k}.
\end{equation}
By $w(z)$ we denote the conformal map inverse to $z(w)$. Remarkably, it can be expressed through the dispersionless tau-function~\eqref{F0} as follows:
\begin{equation}\label{w-tau}
		w(z) = z\exp\left(-\frac12\p_{t_0}^2F - \p_{t_0}D(z)F\right), \qquad D(z) = \sum_{k\geq1}\frac{z^{-k}}{k}\p_{t_k}.
\end{equation}
Besides, $F$ is a potential function for the complimentary set of moments
\begin{equation}
	v_0 = \frac{1}{\pi}\int_D\log|z|^2 \sigma(z,\bar z)d^2z,\qquad v_k = \frac{1}{\pi}\int_D z^k \sigma(z,\bar z)d^2z.
\end{equation}
Namely, one can show that $v_k = \p F/\p t_k$, $k=0,1\dotsc$, and similar relations hold for the complex conjugate moments.

It is convenient to introduce a generating function for all moments $t_k$ and $v_k$. Let $S(z)$ be an analytic continuation of the
function $\p_z U(z,\bar z)$ away from the curve
\begin{equation}\label{S-def}
	S (z ) = \p_z U(z,\bar z), \quad z\in \p D.
\end{equation}
In the case of the quasi-harmonic potential, $U(z,\bar z) = z\bar z$, the function $S(z) = \bar z$, $z\in\p D$, is known as the \textit{Schwarz function} of the curve~\cite{DavisBook}. Below, we will use the same name for the function~\eqref{S-def}, although the correct name should be the generalized Schwarz function. For analytic curves $S(z)$ is holomorphic in some strip-like neighborhood of the curve $ \p D$, and can be represented in the form $S(z) = S_-(z) + S_+(z)$, where $S_-$ is regular in $D^c$, and $S_+$ is regular in $D$. In the neighbor of the contour, the Schwarz function~\eqref{S-def} can be written in terms of Laurent series: 
\begin{equation}\label{S-series}
	S(z) = \sum_{k\geq1}kt_kz^{k-1} + \frac{t_0}{z}+\sum_{k\geq1}v_kz^{-k-1}.
\end{equation}

The Schwarz function plays an important role in the LG problem. In Appendix~\ref{A:S-v} we prove that the following equation holds on the moving boundary of the domain:
\begin{equation}\label{Sv-def}
	\dot S	= 2 \sigma \bar v,
\end{equation}
where the dot stands for the partial derivative w.r.t.\ time, and $\bar v$ denotes complex conjugate of the velocity, $v = v_x + iv_y$, of the boundary of the droplet. In the case of the quasi-harmonic potential, i.e.\ when $\sigma = 1$, we obtain the conventional formulation of the LG problem written in terms of the Schwarz function~\cite{HowisonEJAM}.


\section{Discrete stochastic Laplacian growth}\label{s:growth}


\subsection{The overlap of wave-functions}


In the previous section, we mentioned a possible physical interpretation of $\tau_N$ as the partition function of Coulomb gas in the external potential. Another useful interpretation of the Coulomb gas distribution~\eqref{tau-def} is a coherent state of relativistic electrons in the Quantum Hall regime~\cite{ABWZ02}. From this point of view, we consider $N$ electrons in the plane in a  strong not necessarily uniform magnetic field $B = -\frac12\Delta W$. The exact $N$-particle wave function of electrons that fully occupy the lowest energy level reads:
\begin{equation}\label{psi-def}
	\Psi_N(z_1,\dots,z_N) = \frac{1}{\sqrt{N! \tau_N}}\Delta_N(z)\prod_{j=1}^N e^{\frac{1}{2\hbar} W(z_j,\bar z_j)}.
\end{equation}
The probability to find electrons at positions $\{z_j\}_{j=1}^N$ is given by the joint probability distribution function $|\Psi_N(z_1,\dotsc, z_N)|^2$, i.e.\ the density of the partition function~\eqref{tau-def}. In the semiclassical limit, the wave function~\eqref{psi-def} describes the incompressible electronic droplet. 

We consider the evolution of the electronic droplet under a change in the number of electrons in the droplet. We emphasize that the parameters $\{t_k, \bar t_k\}$ of the external potential~\eqref{W-def} are kept fixed. If one varies the number of electrons, $N\to N + M$ the support of eigenvalues slightly changes its shape (because the process is stochastic) and area. Let $\Psi_{N+ M}$ denotes the $(N+M)$-particle wave-function of the deformed droplet. In order to describe the transition between $N$ and $N+M$ particle states of the system of electrons, we consider the following set of the \textit{overlap functions} on the complex plane:
\begin{multline}\label{overlap-def}
	\psi_{N,M}(z_1,\dotsc,z_{M}) = \\
	=\sqrt{\frac{(N+M)!}{N!}}\int_{\mathbb C}\cdots\int_{\mathbb C} \overline{\Psi_N(z_1',\dotsc,z_N')}\Psi_{N+M}(z_1',\dotsc,z_N',z_1,\dots, z_{M}) d^2z_1'\cdots d^2z_N'.
\end{multline}
Taking into account~\eqref{psi-def} we recast the overlap function~\eqref{overlap-def} in the form:
\begin{multline}\label{overlap-1}
	\psi_{N,M}(z_1,\dotsc,z_{M}) 
	= \frac{1}{N!\sqrt{\tau_N \tau_{N+M}}}\Delta_M(z)\prod_{i=1}^{M}z_i^N e^{\frac{1}{2\hbar}W(z_i,\bar z_i)}\times\\
	\times\int_{\mathbb C}\cdots\int_{\mathbb C} |\Delta_M(z')|^2 \prod_{j=1}^N e^{\frac{1}{\hbar}W(z'_j,\bar z'_j)+\log(1-z'_j/z_i)}d^2z_j'.
\end{multline}

One can expand the logarithmic terms in series in $(z'_j/z_i)$ in the exponent of~\eqref{overlap-1}. It is easy to see, that this series expansion is equivalent to changing the parameters of the potential, $t_k\to t_k-\frac{1}{k}z^{-k}$, which is also known as Miwa transformation. Hence, the $N$-fold integral in~\eqref{overlap-1} can be written as follows:
\begin{equation}\label{tr-1}
	\int_{\mathbb C}\dots\int_{\mathbb C} |\Delta_M(z')|^2  \prod_{j=1}^N\prod_{i=1}^{M}e^{\frac{1}{\hbar}W(z'_j,\bar z'_j)+\log(1-z'_j/z_i)}d^2z_j' = N!\, \tau_N(W+[z_1]+\cdots+[z_{M}]),
\end{equation}
where $W+[z]$ denotes the potential with the shifted parameters $t_k\to t_k-\frac{1}{k}z^{-k}$. By using the operators $D(z)$ introduced in~\eqref{w-tau} one can represent the tau-function with shifted parameters of the potential as the result of the action of the shift operator: 
\begin{equation}\label{tr-2}
	\tau_N(W+[z_1]+\cdots+[z_{M}]) = e^{-\hbar \sum_{j=1}^{M}D(z_j)} \tau_N(W)
\end{equation}
Eqs.~\eqref{tr-1} and~\eqref{tr-2} allow one to recast the overlap function~\eqref{overlap-1} in the form:
\begin{equation}\label{psi-tau}
	\psi_{N,M}(z_1,\dotsc,z_{M}) 
	= \frac{1}{\sqrt{\tau_N \tau_{N+ M}}} \Delta_M(z)\prod_{j=1}^{M}e^{\frac{1}{2\hbar}W(z_j,\bar z_j)}z_j^Ne^{-\hbar D(z_j)}\tau_N,
\end{equation}

As mentioned, in the semiclassical limit (when $N$ is large, $\hbar$ is small, and $N\hbar$ is fixed) the electronic droplet behaves like an incompressible electronic droplet with a well-defined edge. Hence, one can expect that the overlap function~\eqref{psi-tau} can be further simplified in this limit. By taking into account the $\hbar$-expansion of the partition function~\eqref{tau-asymptotic}, and using the Baker-Campbell-Hausdorff formula,
\begin{equation}\label{tr-3}
	e^{-\hbar \sum_j D(z_j)} \tau_N = \tau_N e^{-\frac{1}{\hbar}\sum_j D(z_j)F+\frac12 \sum_{i,j}D(z_i)D(z_j)F},
\end{equation}
we obtain the ratio of the tau-functions, $\tau_N$ and $\tau_{N+M}$, in the r.h.s.\ of Eq.~\eqref{psi-tau}. The leading contribution of $\hbar$-expansion of the free energy $F$ to the ratio reads
\begin{equation}\label{tau/tau}
	\frac{ \tau_{N+M}}{ \tau_N} = (c_p\,\hbar)^\frac{M}{2}\exp\left( \frac{M}{\hbar}\p_{t_0}F + \frac{M^2}{2}\p_{t_0}^2F+O(\hbar)\right),
\end{equation}
where we have used~\eqref{tau-asymptotic} and took into account that $M\hbar\ll1$. It is worth reminding that both tau-functions, $\tau_{N}$ and $\tau_{N+M}$, are determined by the same external potential~\eqref{W-def}, i.e.\ the set of external moments $\{t_k,\bar t_k\}$ is kept fixed during the stochastic growth process. We change the size of matrices only, $N\to N+M$. Hence, only partial derivatives w.r.t.\ the moment $t_0$ appear in~\eqref{tau/tau}.


\subsection{Transition probability in the semiclassical limit}\label{sec:psi-semi}


In this section, we discuss the probability of transition between the $N$ and $(N+M)$- particle states. It can be written as $P(z_1,\dotsc,z_M)\prod_{j=1}^M d^2 z_j$, where the joint probability density function $P(z_1,\dotsc, z_M)$ is given by the amplitude of the overlap function~\eqref{psi-tau} squared, 
\begin{equation}\label{P-def}
	P(z_1,\dotsc,z_M) = \lim_{N\to\infty}|\psi_{N,M}(z_1,\dotsc,z_M) |^2.
\end{equation}
By using~\eqref{tr-3} and~\eqref{tau/tau} we recast $P(z_1,\dotsc, z_M)$ in the form:
\begin{multline}\label{Psi-F}
	P(z_1,\dotsc,z_M) \simeq  (c_p\,\hbar)^{-\frac{M}{2}}|\Delta(z)|^2\times\\
	\times
	\exp\left(-\frac{2}{\hbar}\sum_j\mathcal A(z_j,\bar z_j)-\sum_{i,j}\left[\frac12\p_{t_0}^2-\Re D(z_i)D(z_j)\right] F\right),
\end{multline}
where we introduced the so-called \textit{modified Schwarz potential}:
\begin{equation}\label{A-def}
	\mathcal A(z,\bar z) = \frac{U(z,\bar z)}{2} - \Re \Omega(z),
\end{equation}
and $ \Omega(z)$ is the \textit{generating function}:
\begin{equation}
	\Omega( z ) \equiv \int^z S(\zeta)d\zeta = \sum_{k>0}t_k z^k + t_0\ln z - \left[\frac12\p_{t_0}+D(z)\right]F.
\end{equation}

As discussed, the integrable structure of the problem allows one to express the conformal map, $w:\ D^c\to\mathbb C\setminus \mathbb D$, in terms of the free energy~\eqref{w-tau}. In particular, the following identity holds~\cite{BOOKtau, Z01}:
\begin{equation}\label{w-identity-1}
	\left(-\frac12\p_{t_0}^2+D(z_1)D(z_2)\right)F = \ln\frac{w(z_1)-w(z_2)}{z_1-z_2}.
\end{equation}
By taking into account~\eqref{w-identity-1} we recast~\eqref{Psi-F} in the form:
\begin{equation}\label{psi-result}
	P(z_1,\dotsc,z_M) \simeq (c_p\,\hbar)^{-\frac{M}{2}}
	\prod_{1\leq n<m\leq M}|w(z_n)-w(z_m)|^2\prod_{j=1}^M|w'(z_j)|e^{-\frac{2}{\hbar}\mathcal A(z_j,\bar z_j)}.
\end{equation}
Note, that this formula is valid for any $M\geq1$. In the case when we add only one eigenvalue to the droplet at the point $z\in D^c$, i.e.\ $M=1$,  the probability density function takes the form~\cite{ABWZ02}:
\begin{equation}\label{psi-1}
	P(z) \simeq \frac{1}{\sqrt{2\pi^3 \hbar}}|w'(z)|e^{-\frac2\hbar \mathcal A(z,\bar z)}.
\end{equation}
The probability distribution of adding $M$ eigenvalues to the droplet~\eqref{psi-result} is given by the product of the one-point probabilities~\eqref{psi-1} times the Vandermonde determinant squared $|\Delta(w(z))|^2$.

We can better understand the behavior of the function $P(z_1,\dotsc,z_M) $ by studying the properties of the modified Schwarz potential $\mathcal A$ near the boundary. This potential reaches its minimum on the edge of the droplet, $\mathcal A(z,\bar z) = 0$ when $z\in \p D$. Also, from the definition of the Schwarz function~\eqref{S-def} it follows $\p\mathcal A(z,\bar z) = 0$ $(z\in\p D)$ when $z\in\p D$. In Appendix~\ref{A:A-expansion} we prove that
\begin{equation}\label{A-expansion}
	\mathcal A(z+\delta_n z,\overline{z+\delta_n z}) = \sigma(z,\bar z)|\delta_n z|^2 + O\left((\delta_n z)^3\right), \qquad z\in \p D,
\end{equation}
where $|\delta_n z|$ is the displacement of the boundary in the normal direction. From~\eqref{A-expansion} we conclude that the function~\eqref{psi-result} is localized in the vicinity of $\p D$, where the amplitude has a sharp maximum. We see that the width of the Gaussian distribution depends on the point of the curve through the function $\sigma$. In particular, the Gaussian function tends to the delta function with the support on the curve as $\hbar\to0$.


\subsection{The semiclassical limit of the overlap function}


Starting from this section we study the double scaling limit of the growth process of normal matrices when both $N$ and $M$ are large, and $1\ll M\ll N$. By $D_{t}$ and $D_{t + \delta t}$ we denote the domains constituted of $N$ and $(N+M)$ eigenvalues correspondingly. It is convenient to introduce the density of eigenvalues $\rho(z) = \sum_{i=1}^M \delta^{(2)}(z-z_i)$, which becomes a smooth function in the large $M$ limit. In this limit the joint probability density function~\eqref{P-def} becomes a functional on the density of eigenvalues,
\begin{equation}\label{P-def-limit}
	P[\rho] = \lim_{M\to \infty} P(z_1,\dotsc,z_M).
\end{equation}
In order to elaborate the large $M$ limit of the growth probability we recast~\eqref{psi-result} in terms of the density function. In the leading order in $\hbar$ we have:
\begin{multline}\label{Prho}
	P[\rho] \simeq (c_p\,\hbar)^{-\frac{M}{2}}
	\exp\left\{\frac{1}{\hbar^2}\left(\iint \rho(z)\log|w(z)-w(\zeta)|\rho(\zeta)d^2z d^2\zeta - \right.\right.\\
	\left.\left.- 2\int \rho(z)\mathcal A(z,\bar z)d^2z\right)\right\}.
\end{multline}
As explained at the end of Sec.~\ref{sec:psi-semi}, the eigenvalues $\{z_j\}_{j=1}^M$ form a thin layer at the boundary of the droplet $\p D_t$, so that $\rho(z)$ is a continuous function inside the layer $\delta D  = D_{t+\delta t}\setminus D_{t}$ normalized as follows: $\int_{\delta D} \rho(z)d^2z = M\hbar$. The most probable distribution of eigenvalues is determined by the maximum of the functional~\eqref{Prho}. The saddle point condition, $ \delta P[\rho]/\delta \rho = 0$, yields the integral equation for the mean density:
\begin{equation}
	\int\log|w(z)-w(\zeta)|\rho(\zeta)d^2\zeta - \mathcal A(z,\bar z) + \frac{\lambda}{2} = 0,
\end{equation}
where $\lambda$ is a Lagrange multiplier, which is required in order to satisfy the normalization constraint $\int \rho(z)dz^2 = M\hbar$. Upon taking the Laplacian of this equation, we see that the ``classical'' density which maximizes the probability equals the background charge $\rho_{cl}(z) = \sigma(z,\bar z)$ for $z\in \delta D$, and $\rho_{cl}(z) = 0$ otherwise. The eigenvalues are localized in the thin layer $\delta D$ at the boundary of the domain $D$. We can recast the functional on the density~\eqref{Prho} in the functional on the shape of the layer provided that $\rho = \rho_{cl}\equiv \sigma$ inside the layer.

First, we rewrite the functional~\eqref{Prho} in the form:
\begin{multline}\label{Prho-deltaD}
	P[\delta D] \simeq  (c_p\,\hbar)^{-\frac{M}{2}}
	\exp\left\{\frac{1}{\hbar^2}\left(\int_{\delta D}\int_{\delta D} \sigma(z, \bar z)\log|w(z)-w(\zeta)|\sigma(\zeta,\bar\zeta)d^2z d^2\zeta - \right.\right.\\
	\left.\left.- 2\int_{\delta D} \sigma(z,\bar z)\mathcal A(z,\bar z)d^2z\right)\right\}.
\end{multline}
In Appendix~\ref{A:Stokes} we prove Stokes' theorem which allows us to recast the integral over the layer $\delta D$ in the contour integral along the inner boundary of the layer $\p D$. Namely, we show that the probability density~\eqref{Prho-deltaD} can be rewritten as follows:
\begin{equation}\label{psi2-deltaS}
	P[\delta S] \simeq  (c_p\,\hbar)^{-\frac{M}{2}}
	\exp\left\{\frac{1}{\hbar^2}\Re\left(\oint_{ \p D}\oint_{ \p D} \delta S(z)\log(w(z) - w(\zeta)) \delta S(\zeta)dz d\zeta\right)\right\},
\end{equation}
where $ \delta S(z)= S_{t+ \delta t}(z) - S_{t}(z)$ denotes the variation of the Schwarz function~\eqref{S-def} under the variation of the domain $D_{t}\to D_{t+\delta t}$. The Schwarz functions $S_t$ and $S_{t+\delta t}$ at $z\in \p D_t$ are given by the Laurent series~\eqref{S-series}, where the series coefficients are the harmonic moments of the domains $D_t$ and $D_{t+ \delta t}$ correspondingly. Hence, the variation of the Schwarz function can be written in terms of variations of coefficients of the Laurent series:
\begin{equation}\label{deltaS-series}
	\delta S(z) = \sum_{k\geq1}k \delta t_kz^{k-1} + \frac{\delta t_0}{z}+\sum_{k\geq1}\delta v_kz^{-k-1}.
\end{equation}

It is more convenient to describe the interface dynamics in terms of the normal interface velocity $v_n(z)$ rather than in terms of the time-dependent Schwarz function $S(z)$. As explained, in LG the variation of the Schwarz function is connected to the interface velocity by Eq.~\eqref{Sv-def}. By taking into account that $dz = in(z)|dz|$, where $n(z)=-i \tau(z)$ is the unit normal vector to $\p D$ at the point $z$ orthogonal to the unit tangent $\tau(z)$, we recast the contour integrals in~\eqref{psi2-deltaS} into the line integrals along the boundary of the domain. In the case of zero surface tension, pressure is constant along the boundary and so its gradient is orthogonal to it. In the complex notation, this constraint can be written as follows: $\Re(\tau\bar v) = 0$. Therefore, from Eq.~\eqref{Sv-def} we conclude that the product of the functions $\dot S(z)n(z)$ is real. Hence, we can rewrite~\eqref{psi2-deltaS} as follows:
\begin{equation}\label{psi2-deltah}
	P[h] \simeq (c_p\,\hbar)^{-\frac{M}{2}}\left\{\frac{1}{\hbar^2}\oint_{ \p D}\oint_{ \p D} h(z)\sigma(z,\bar z)\log|w(z) - w(\zeta)| \sigma(\zeta,\bar \zeta) h(\zeta) |dz| |d\zeta|\right\},
\end{equation}
where $ h(z) = v_n(z)\delta t$ is a width of the layer $\delta D$ at the point $z\in \p D$, and $v_n = \bar v n$ is the normal interface velocity. One can refer to the functional~\eqref{psi2-deltah} as the probability density of random variables $h(z)$ for the stochastic LG problem in the semiclassical limit. It specifies the probabilities of outcomes of the growth process, $N\to N+M$, of the support of eigenvalues of normal random matrices.

In concluding this section, let us draw a connection between the growth process of normal random matrices considered in this paper and the stochastic LG model introduced in~\cite{AM16}. Stochastic LG is defined as the discrete aggregation process in the plane where instead of one particle the source emits $M\gg1$ uncorrelated particles of area $\hbar$ per time unit $\delta t$. The particles quickly diffuse until they stick to the boundary of the growing domain forming on its surface an external layer $\delta D$ of area $M\hbar$. The probability of a particular growth step is determined by the multinomial distribution. Remarkably, the multinomial distribution function of the layer in stochastic LG reduces to~\eqref{psi2-deltah} in the semiclassical limit provided that $\sigma(z,\bar z)=1$~\cite{AM17a}.

\subsection{The ``classical'' dynamics reproduces LG}


Let us prove that the most probable interface dynamics, i.e.\ the ``classical'' displacement of the boundary, $h^{cl}$, maximizing the probability density~\eqref{psi2-deltah} is similar to standard LG. The extremum condition, $ \delta P[h]/\delta h = 0$, for the functional $P[h]$ yields the integral equation, which can be solved in terms of the Green's function,
\begin{equation}\label{Sc-def}
	h_{cl}(z) = -\frac{\epsilon}{2\pi \sigma(z,\bar z)}\p_n G(z,\infty).
\end{equation}
Here $\epsilon = M\hbar $ is the area of the layer $\delta D$, and $\p_n G(z,\infty)$ denotes the normal derivative of the Green's function $G(z,z')$ of the Dirichlet problem taken at the boundary with the normal vector being directed to $D^c$.

We briefly recall the main properties of the Green's function. By definition $G(z,z')$ in $D^c$ is a harmonic function except the point $z=z'$, where $G(z,z')$ diverges logarithmically: $G(z,z')\to\log|z-z'|$ as $z\to z'$. The Green function is symmetric in $z,\ z'$ and $G(z,z') = 0$ at the boundary. It can be written explicitly in terms of the conformal map $w(z)$ from $D^c$ onto the exterior of the unit disk,
\begin{equation}\label{G-def}
	G(z,z') = \log\left|\frac{w(z) - w(z')}{1 - w(z)\overline{w(z')}}\right|.
\end{equation}

From~\eqref{G-def} we obtain that $G(z,\infty) = -\log|w(z)|$, and $\p_n G(z,\infty) = -|w'(z)|$ on $\p D$. The normal interface velocity $v_n(z)$ is defined as the normal displacement of the boundary $h(z)$ per time instant $\delta t$, namely, $v_n^{cl} = \lim_{\delta t\to 0} h_{cl}/\delta t$. Rewriting Eq.~\eqref{Sc-def} in terms of the conformal map $w(z)$ we obtain
\begin{equation}\label{Darcy}
	v^{cl}_n(z) = \frac{\dot \epsilon}{2\pi\sigma(z,\bar z)}|w'(z)|,\quad z\in\p D,
\end{equation}
where $\dot \epsilon = \epsilon/\delta t$ is the growth rate (area growth per time unit). In the case of the uniform background charge distribution, $\sigma(z,\bar z) = 1$, we recover the standard LG dynamics when the normal interface velocity is proportional to the pressure gradient at the boundary: $v_n(z)\propto |w'(z)|$. This is Darcy's law for the dynamics of the interface between viscous and non-viscous fluids confined in the Hele-Shaw cell, assuming vanishing surface tension at the interface~\cite{BensimonRMP, EtingofBook}. For a non-constant $\sigma$ we have LG in the non-uniform background.

The ``nonclassical'' growth scenarios, $h\neq h^{cl}$, result in deviations of the interface dynamics from standard LG. One can interpret these deviations as stochastic fluctuations of the interface (or noise) distributed following the probability density~\eqref{psi2-deltah}. It is easy to see that the density~\eqref{psi2-deltah} is similar to the probability density of $M$ eigenvalues of the circular unitary ensemble (CUE). Hence, the $n$-point correlation functions of fluctuations at the boundary of the growing domain can be determined explicitly. We will address the correlation functions in future publications.


\section{Universality of fluctuations}\label{s:universality}


We will refer to the  normalization factor in~\eqref{psi-result},
\begin{equation}\label{Z-fluctuations}
	\mathcal Z = \int_{D^c}\cdots\int_{D^c} |\Delta(w(z))|^2\prod_{j=1}^M|w'(z_j)|e^{-\frac{2}{\hbar}\mathcal A(z_j,\bar z_j)}\frac{d^2z_j}{\hbar^{1/2}},
\end{equation}
as to the partition function of the fluctuations. Let us discuss first the case of the quasiharmonic potentials with $\sigma(z,\bar z) = 1$. Using~\eqref{A-expansion} one can replace the integrals over the layer by the integrals along the boundary of the domain $\p D$. Further, by parametrizing the boundary,  $z = z(e^{i\phi})$, by the angle $\phi$ in the $w$ plane, and changing the integration variables, $|dz(w)| = |z'(w)||dw|$, we compute the integrals in~\eqref{Z-fluctuations} explicitly:
\begin{equation}\label{Z-quasi}
	\mathcal Z_\text{quasi-harm} = (2\pi^3)^{M/2}.
\end{equation}
Note, that this result immediately follows from the normalization condition for the probability density~\eqref{psi-result}, namely, $\int P(z_1,\dotsc, z_M)\prod d^2z_j = 1$, where we took into account that $c_p = 2\pi^3$ in the case of the quasi-harmonic potentials (see discussion below Eq.~\eqref{tau-asymptotic}). In the case of non-uniform background charge distribution, $\sigma\neq 1$, the partition function of fluctuations reads:
\begin{equation}
	\mathcal Z_p = (c_p)^{M/2}.
\end{equation}
It is important, that $c_p$ is a numerical constant. Therefore, the partition function depends on the number of added eigenvalues $M$ only.

Remarkably, the partition function for fluctuations~\eqref{Z-fluctuations} is \textit{universal}. In contrast to the tau-function of the domain~\eqref{tau-def} the partition function of the layer $\mathcal Z$ does not depend on the domain $D$ and, correspondingly, on the moments $\{t_0,t_k,\bar t_k\}$ of the external potential~\eqref{W-def}. It only depends on the area of the layer $M\hbar$ in the semiclassical limit.


\section{Conclusion}

 
We have studied stochastic Laplacian growth in the framework of normal random matrices and show that in the semiclassical limit, a change in the size of matrices can be interpreted as the evolution of the boundary of the support of eigenvalues. The growth probability was defined as the amplitude of the matrix element for $N$ and $(N+M)$-particle states, and shown to satisfy the Gibbs-Boltzmann statistics. When $M=1$ the growth follows DLA rules as shown in Ref.~\cite{ABWZ02}. We generalized this result for the case $M>1$ and obtained the joint probability distribution of adding $M$ eigenvalues in the semiclassical limit.

When $M$ becomes large, $1\ll M\ll N$, the added eigenvalues form an external layer at the boundary of the initial domain. In this case, the probability density~\eqref{psi2-deltah} (upon changing the variables from the $z$ plane to the $w$ plane) is similar to the probability distribution of $M$ eigenvalues in CUE. The interface dynamics is stochastic, and the most probable growth process is equivalent to deterministic LG. Random fluctuations deflect moving fronts from the classical growth scenario. The probability density function of fluctuations can be obtained from~\eqref{psi2-deltah} by expanding $h=h^{cl}+\delta h$.

It is worth mentioning that we studied the growth of normal random matrices in non-uniform background charge distribution $\sigma(z,\bar z)$ in the complex plane. Our main intention was the case when $\sigma(z,\bar z) = \sigma(|z|)$ is the axially symmetric function in $|z|$. In this case, the problem with non-uniform $\sigma$ can be mapped to another LG problem with uniform $\sigma$ but with different boundary conditions. Let us mention some important examples: the problems with $\sigma(|z|) = |z|^{2 \alpha-2}$ and $\sigma(|z|) = 1/|z|^{2}$ can be mapped to the growth problems in the wedge with the opening angle $2\pi \alpha$ and in a channel with periodic boundary conditions, correspondingly. Hence, our results are rather general and can be used to study stochastic growth not only in the plane but also in the channel and the wedge.

The universality of stochastic LG is another important result. We have shown that the partition function of fluctuations (or, the layer) does not depend on the shape of the initial domain. This feature of the model agrees well with the experimental observations: the patterns typically observed in the Hele-Shaw cell are scale-invariant multi-branched fractals, which are universal (independent of local geometric details). The universality of stochastic LG might suggest that all relevant statistical properties of growing patterns (e.g., fractal dimension) can be determined from the partition function of fluctuations.


\section*{Acknowledgment}
The work is supported by the Russian Science Foundation grant 19-71-30002.


\appendix


\section{LG in the non-uniform background metric in terms of the Schwarz function}\label{A:S-v}


Suppose that $z \in \p D_t$. At the time instant $t + d t$ the point $z+vdt$ belongs to the new boundary $\p D_{t+dt}$, and Eq.~\eqref{S-def} takes the form $S(z+vdt,t+dt) = \p_zU(z+vdt,\overline{z+vdt})$. Therefore, we obtain the following relation:
\begin{equation}\label{barv-1}
	\sigma \bar v = \dot S + (S' - U'')v,
\end{equation}
which holds on the boundary $z\in\p D_t$. Here, the dot and prime denote partial derivative w.r.t.\ time and $z$ correspondingly. In the case of zero surface tension pressure is constant along the boundary, and so its gradient is orthogonal to it. In the complex notation, we have
\begin{equation}\label{barv-tau}
	\Re(\bar v \tau)=0,
\end{equation}
where $\tau(z)$ is the unit tangent at $z\in\p D_t$. Let us note that the line element $|dz|$ along the curve $\p D$ can be written as
\begin{equation}
	|dz| = \sqrt{dz d\bar z} = \sqrt{\frac{S'-U''}{\sigma}}dz,
\end{equation}
where we used Eq.~\eqref{dz-dbarz}. It follows then that the unit tangent vector to $\p D$ at a point $z$ is $\tau = dz/|dz|=\sqrt{\sigma/(S'-U'')}$. Therefore, from~\eqref{barv-tau} we obtain
\begin{equation}\label{barv-2}
	\bar v + \frac{S'-U''}{\sigma} v = 0.
\end{equation}
Using equation~\eqref{barv-1} and~\eqref{barv-2} we then find the desired relation between the variation of the Schwarz function and the velocity of the boundary in LG: 
\begin{equation}
    \dot S = 2 \sigma \bar v.
\end{equation}


\section{Expansion of $\mathcal A(z,\bar z)$ near the boundary}\label{A:A-expansion}


We consider the behavior of the modified Schwarz potential $\mathcal A$ near the edge of the domain. It is easy to see that the series expansion of $\mathcal A(z+\delta z)$ near the edge starts with the second order in $\delta z$. By taking into account~\eqref{A-def} we can find the second-order variation of the potential:
\begin{equation}
	2\mathcal A(z+\delta z) = \sigma(z,\bar z)|\delta z|^2 - \Re\left((S'(z)-U''(z,\bar z))(\delta z)^2\right)+O((\delta z)^3),
\end{equation}
where prime denotes the partial derivatives w.r.t.\ $z$. We can represent the variation $\delta z = \delta_n z + \delta_\tau z$ as the sum of the normal and tangential variations of the boundary, $\delta_n z$ and $\delta_\tau z$, corresponding. From the definition of the Schwarz function~\eqref{S-def} it follows
\begin{equation}\label{dz-dbarz}
	\sigma d\bar z = (S'-U'')dz.	
\end{equation}
By taking account of this equality we obtain:
\begin{equation}
	|\delta_{\tau}z| = \sqrt{\delta_\tau z \delta_\tau\bar z} = \sqrt{\frac{\delta_{\tau} \bar z}{\delta_{\tau} z}}\delta_\tau z = \sqrt{\frac{S'-U''}{\sigma}}\delta_\tau z, 
\end{equation}
Besides, $\delta_n z = \mp |\delta_n z/\delta_\tau z|\delta_\tau z$, where the upper (lower) sign should be taken for the outward (inward) deviation. Using these relations we readily obtain the leading contribution to the expansion of $\mathcal A(z,\bar z)$ in the vicinity of the $\p D$:
\begin{equation}
	\mathcal A(z+\delta_n z,\overline{z+\delta_n z}) = \sigma(z,\bar z)|\delta_n z|^2 + O\left((\delta_n z)^3\right), \qquad z\in \p D.
\end{equation}


\section{Stokes' theorem}\label{A:Stokes}


For any smooth complex-valued function $g$ defined on a domain $D$ and its boundary $\p D$, Stokes' theorem can be written in the following form:
\begin{equation}\label{Stokes-def}
	\int_{D}\bar\p g(z,\bar z)d^2z = \frac{1}{2i}\oint_{\p D} g(z,\bar z) dz,
\end{equation}
where $d^2z := dxdy$ is the area element. Let us prove Stokes' theorem for the integral in the exponent in the r.h.s. of~\eqref{Prho-deltaD}:
\begin{multline}\label{Stokes}
	\Re\int_{\delta D}\int_{\delta D} \sigma(z, \bar z)\log(w(z)-w(\zeta))\sigma(\zeta,\bar\zeta)d^2z d^2\zeta - 2\int_{\delta D} \sigma(z,\bar z)\mathcal A(z,\bar z)d^2z = \\
	= \Re\oint_{\p D}\oint_{\p D} \delta S(z)\log(w(z)-w(\zeta))\delta S(\zeta) \frac{dz d\zeta}{(2i)^2},
\end{multline}
provided that $\delta D=D_{t+\delta t}\setminus D_t$ is the layer, $\p D$ is the inner boundary of the layer, and $\delta S(z) = S_{t+\delta t}(z) - S_t(z)$ is the variation of the Schwarz function w.r.t.\ the variation of the domain $D_{t} \to D_{t+\delta t}$. Let $I$ denotes the l.h.s.\ of~\eqref{Stokes}. We represent $I$ as a sum of two terms $I=I_1+I_2$, where $I_1$ and $I_2$ are given by
\begin{equation}\label{I1I2}
	\begin{gathered}
		I_1 = \Re\int_{\delta D}\int_{\delta D} \sigma(z, \bar z)\log\left(\frac{w(z)-w(\zeta)}{z-\zeta}\right)\sigma(\zeta,\bar\zeta)d^2z d^2\zeta,\\
		I_2 = \Re\int_{\delta D}\int_{\delta D} \sigma(z, \bar z)\log(z-z')\sigma(\zeta,\bar\zeta)d^2z d^2\zeta - 2\int_{\delta D} \sigma(z,\bar z)\mathcal A(z,\bar z)d^2z.
	\end{gathered}
\end{equation} 
The transformation of the first integral $I_1$ is straightforward because the integrand is regular inside the layer $\delta D$. By applying Stokes' theorem~\eqref{Stokes-def} twice we readily obtain
\begin{equation}\label{I_1-def}
	I_1 = \Re\oint_{\p D}\oint_{\p D}\delta S(z)\log\left(\frac{w(z)-w(\zeta)}{z-\zeta}\right)\delta S(\zeta)\frac{dz d\zeta}{(2i)^2}.
\end{equation}
Here $\delta S(z) = S_{t+\delta t}(z) - S_t(z)$ is the variation of the Schwarz function at $z\in\p D$. Besides, we changed the integration along the outer boundary of the layer by the integration along the inner boundary of the layer, because the integrand is a regular function in $\delta D$. 

The second integral $I_2$ requires close attention due to the logarithmic branch cut in the integrand. Let us prove that this integral is equal to
\begin{equation}\label{I2-def}
    I_2 = \Re\oint_{\p D}\oint_{\p D}\delta S(z)\log(z-\zeta)\delta S(\zeta)\frac{dz d\zeta}{(2i)^2}.
\end{equation}
It is convenient to start with the integral~\eqref{I2-def} and recast it to the integral over the layer using Stokes' theorem. We rewrite $I_2$ in the form $I_2 = \Re \int_{\p D} J(z)\delta S(z)\frac{dz}{2\pi i}$, and consider the auxiliary integral $J(z)$ along the boundary
\begin{equation}\label{app:J-def}
	J(z) = \oint_{\p D}\frac{d\zeta}{2\pi i}\delta S(\zeta)\log(z-\zeta) = \oint_{\p D}\frac{d\zeta}{2\pi i}(S_{t + \delta t}(\zeta)-S_t(\zeta))\log(z-\zeta),
\end{equation}
where $z\in \delta D$. By using Stokes' theorem we recast $J(z)$ in the integral over the layer $\delta D$. Upon transformation of the contour $\p D (=\p D_t)\to\p D_{t+\delta t}$ in the first integral in the r.h.s.\ in~\eqref{app:J-def} we obtain the additional term $-\int_{[z,z_0]}S_{t+\delta t}(\zeta)d\zeta$, where $[z,z_0]$ is a cut inside the layer with the endpoints $z\in \delta D$ and $z_0\in\p D$. We can add and subtract the following integral to $J(z)$:
\begin{equation}
	J_\text{cut}(z,z_0) =  \frac{1}{2\pi i}\left(\int_{[z,z_0]^+}-\int_{[z,z_0]^-}\right)S_\text{cut}(\zeta)\log(z-\zeta)d\zeta = - \int_{[z,z_0]}S_\text{cut}(\zeta)d\zeta,
\end{equation}  
where $S_\text{cut}(z)$ is the Schwarz function of the cut, and $[z,z_0]^\pm$ are the upper and lower edges of the cut correspondingly. Taking account of Stokes' theorem we recast the sum of the integrals $J(z)+J_\text{cut}(z)-J_\text{cut}(z)$ in the integral over the layer:
\begin{equation}\label{J-tr}
    J(z) = \int_{\delta D}\sigma(z,\bar z)\log(z-\zeta)d^2\zeta - \int_{[z,z_0]} (S_t(\zeta)-S_\text{cut}(\zeta))d\zeta.
\end{equation}
Recall that $\int^z S(\zeta)d\zeta = \Omega(z)$, and $\Re \Omega(z) = |z|^2/2$ if $z\in\p D$. Since $z_0\in\p D$ the contributions of $z_0$ to $\Re J(z)$ in the r.h.s.\ of~\eqref{J-tr} cancel each other. By applying the Stokes' theorem to second integral $I_2 = \Re\int_{\p D} J(z)\delta S(z)\frac{dz}{2\pi i}$ we obtain exactly the integral $I_2$ from Eq.~\eqref{I1I2}. This completes the proof.


\bibliographystyle{ieeetr}
\bibliography{biblio}{}
\end{document}